\documentclass{article}

\usepackage{arxiv}

\usepackage[utf8]{inputenc} % allow utf-8 input
\usepackage[T1]{fontenc}    % use 8-bit T1 fonts
\usepackage{hyperref}       % hyperlinks
\usepackage{url}            % simple URL typesetting
\usepackage{booktabs}       % professional-quality tables
\usepackage{amsfonts}       % blackboard math symbols
\usepackage{nicefrac}       % compact symbols for 1/2, etc.
\usepackage{microtype}      % microtypography
\usepackage{lipsum}		% Can be removed after putting your text content
\usepackage{graphicx}
\usepackage{amsmath,amsthm,amssymb}
\usepackage[numbers]{natbib}

\def\det{\text{det}}

\title{SEIAR model with asymptomatic cohort and consequences to efficiency of quarantine government measures in COVID-19 outbreak}

\author{ \href{https://orcid.org/0000-0002-9027-4333}{\includegraphics[scale=0.06]{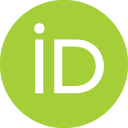}\hspace{1mm}Lenka P\v{r}ibylov\'{a}}\thanks{corresponding author} \\
	Department of Mathematics and Statistics\\
	Masaryk University\\
	 Kotl\'a\v rsk\'a 2,
Brno, 611 37, Czech Republic \\
	\texttt{pribylova@math.muni.cz} \\
	\And
	 \href{https://orcid.org/0000-0001-8476-7740}{\includegraphics[scale=0.06]{orcid.png}\hspace{1mm}Veronika Hajnov\'{a}} \\
	Department of Mathematics and Statistics\\
	Masaryk University\\
	 Kotl\'a\v rsk\'a 2,
Brno, 611 37, Czech Republic \\
	\texttt{hajnova@math.muni.cz} \\
}

\begin{document}
\maketitle

\begin{abstract}
We present a compartmental SEIAR model of epidemic spread as a generalization of the SEIR model. We believe that the asymptomatic infectious cohort is an omitted part of the understanding of the epidemic dynamics of disease COVID-19. We introduce and derive the basic reproduction number as the weighted arithmetic mean of the basic reproduction numbers of the symptomatic and asymptomatic cohorts. Since the asymptomatic subjects people are not detected, they can spread the disease much longer, and this increases the COVID-19 $R_0$ up to around 9. We show that European epidemic outbreaks in various European countries correspond to the simulations with commonly used parameters based on clinical characteristics of the disease COVID-19, but $R_0$ is around three times bigger if the asymptomatic cohort is taken into account. Many voices in the academic world are drawing attention to the asymptomatic group of infectious subjects at present. We are convinced that the asymptomatic cohort plays a crucial role in the spread of the COVID-19 disease, and it has to be understood during government measures.
\end{abstract}

% keywords can be removed

\section{Introduction}
We started modeling the Wuhan COVID-19 outbreak with the standard SEIR model with WHO premised basic reproduction number $R_0$ between 2 and 3, the incubation period around 5 days, and the serial time around a week using GLEAMviz network simulator that includes populations, traffic, and measures. Soon we realized, the simulation predicts outbreaks in European countries too late. The outbreaks fitted better for $R_0$ much higher than it is generally considered. Afterwards, serious outbreaks in Italy, Spain, UK, and the US happened very quickly. The paper \cite{Science} appeared in Science, estimating 86\% of undocumented infectious (published 16.3.2020). Padova University recently informed about an experiment in town V\`o, where areal testing showed that the majority of the positive 3\% of citizens were asymptomatic at the beginning of the outbreak. We made a hypothesis that the asymptomatic cohort can play a crucial role. This could also be a clue to understanding why early enough closing school policy contributes to slow down the outbreak. We propose a model that counts in with this asymptomatic infectious cohort, and we derive its basic reproduction number $R_0$.  We proved that the basic reproduction number is a weighted average of reproduction numbers of both the symptomatic and asymptomatic cohorts, which seems to be very intuitive. On the other hand, this implies that in case of a disease with the majority of asymptomatic cases, this basic reproduction number $R_0$ is much higher than the one gained from standard estimates based only on the symptomatic cohort. This can explain the "invisible" and fast increase in COVID-19 outbreaks since it also acts through the asymptomatic cohort. There are already studies that estimate $R_0$ higher than WHO proposed (for example, Diamond Princess estimate 14.8 at \cite{Princess}). Probably we can see just the tip of the iceberg. The consequences are very important since it gives a much shorter time to governments. Late measures have almost no effect.

\section{SEIAR model and the basic reproduction number}
\label{sec:SEIAR}

We propose a simple model that generalizes the SEIR model commonly used for virus disease modeling. The total population is partitioned into the following compartments: $S$, Susceptible; $E$, Exposed; $I$, Infected (symptomatic infected); $A$, Asymptomatic (asymptomatic infected); $R$, Removed (healed or dead, no more infectious).

\begin{eqnarray}
\label{S}
\dot S & = & - \beta S (I+A),\\
\label{E}
\dot E & = &  \beta S (I+A) - \gamma E,\\
\label{I}
\dot I & = &  \gamma p E - \mu_1 I,\\
\label{A}
\dot A & = &  \gamma (1-p) E - \mu_2 A,\\
\label{R}
\dot R & = & \mu_1 I + \mu_2 A, 
\end{eqnarray}
where parameter $\beta$ denotes the transmission rate (i.e., the probability of disease transmission in a single contact times the average number of contacts per person) due to contacts between a Susceptible subject and an Infected or an Asymptomatic subject\footnote{The model can be improved by incorporating different transmission rates for cohorts $I$ and $A$ as $\beta_I$ and $\beta_A$, respectively. It is difficult to compare them since, on one side, people tend to avoid contact with subjects showing symptoms, but on the other side, an asymptomatic subject is less infectious more probably. Due to this dichotomy, we use one the mean $\beta$.}. Parameter $\beta$ can be modified by quarantine government measures that increase social distancing as closing schools, remote working, using masks, or similar. Parameter $\gamma$ usually denotes the probability rate at which the Exposed subject develops clinically relevant symptoms. The period $1/\gamma$ (days$^{-1}$) is called an incubation period. In the case of COVID-19, it is known that an Exposed subject is infectious one or two days before developing symptoms, and so we will define the Exposed compartment as a latent non-infectious compartment. Due to this, we assume that the period $1/\gamma$ is shorter than usually estimated (see \cite{Lombardy}, \cite{NEJM}). We assume that every Exposed subject becomes infectious, but subjects that enter $I$ cohort later develop symptoms with probability $p$ and those who not enter $A$ cohort with probability $1-p$. This assumption is based on recent studies \cite{Science} and experiments in the Italian town V\`o.  Parameters $\mu_1$ or $\mu_2$, respectively, denote the remove rate, so $\tfrac1{\mu_1}$ is the average period to isolation for Infected symptomatic subjects, and $\tfrac1{\mu_2}$ is the average recovery period for Asymptomatic subjects. We assume that $\tfrac1{\mu_2} >\tfrac1{\mu_1}$, since the COVID-19 patients are treated or isolated very quickly after developing symptoms.
We omit the probability rate of becoming susceptible again after recovery, although this is not evident (mainly for Asymptomatic subjects). We assume $S+I+E+A+R=1$ (a non-dimensionalized model with a constant incidence rate).

\subsection{Unstability of the non-epidemic equilibrium -- outbreak}

The non-epidemic equilibrium of the system \eqref{S}, \eqref{E}, \eqref{I},\eqref{A} (we can separate independent equation \eqref{R}) is obviously $(1,0,0,0)$. The Jacobian linearization matrix of the system \eqref{S}, \eqref{E}, \eqref{I},\eqref{A} is
\begin{equation*}
J=\begin{pmatrix}
-\beta(I+A) & 0 & -\beta S &  -\beta S \\
\beta(I+A) & - \gamma & \beta S &  \beta S \\
0 & \gamma p & -\mu_1 & 0 \\
0 & \gamma (1-p) & 0 & -\mu_2 \\
\end{pmatrix}
\end{equation*}
and in the non-epidemic equilibrium it is
\begin{equation} \label{jacobian}
J(1,0,0,0)=\begin{pmatrix}
0 & 0 & -\beta  &  -\beta  \\
0 & - \gamma & \beta  &  \beta  \\
0 & \gamma p & -\mu_1 & 0 \\
0 & \gamma (1-p) & 0 & -\mu_2 \\
\end{pmatrix}. 
\end{equation}
The Jacobian matrix \eqref{jacobian} has one zero eigenvalue. The non-epidemic equilibrium loses stability as any eigenvalue of a submatrix
\begin{equation*} 
A=\begin{pmatrix}
 - \gamma & \beta  &  \beta  \\
 \gamma p & -\mu_1 & 0 \\
 \gamma (1-p) & 0 & -\mu_2 
\end{pmatrix}
\end{equation*}
 crosses imaginary axes and has positive real part. The characteristic polynomial of the matrix $A$ is
\begin{equation*}
p(\lambda)={\lambda}^{3}+ \left( \gamma+\mu_{{1}}+\mu_{{2}} \right) {\lambda}^{2}
+ \left( -\beta\,\gamma+\gamma\,\mu_{{1}}+\gamma\,\mu_{{2}}+\mu_{{1}}
\mu_{{2}} \right) \lambda+\beta\,\gamma\,p\mu_{{1}}-\beta\,\gamma\,p
\mu_{{2}}-\beta\,\gamma\,\mu_{{1}}+\gamma\,\mu_{{1}}\mu_{{2}}
\end{equation*} 
and Routh-Hurwitz criterion 
\begin{eqnarray}
\label{1}
\ \gamma+\mu_{{1}}+\mu_{{2}} & >& 0,\\
\label{2}
-\det A = \beta\,\gamma\,p\mu_{{1}}-\beta\,\gamma\,p
\mu_{{2}}-\beta\,\gamma\,\mu_{{1}}+\gamma\,\mu_{{1}}\mu_{{2}} & > & 0,\\
\label{3}
\left( \gamma+\mu_{{1}}+\mu_{{2}} \right) \left( -\beta\,\gamma+\gamma\,\mu_{{1}}+\gamma\,\mu_{{2}}+\mu_{{1}}
\mu_{{2}} \right) - \left( \beta\,\gamma\,p\mu_{{1}}-\beta\,\gamma\,p
\mu_{{2}}-\beta\,\gamma\,\mu_{{1}}+\gamma\,\mu_{{1}}\mu_{{2}} \right)& > & 0,
\end{eqnarray}
implies negative real parts of all the eigenvalues of the matrix $A$.
The condition \eqref{1} is satisfied always. Violation of the condition \eqref{2} implies that at least one eigenvalue has positive real part. The condition \eqref{2} can be equivalently rewritten as
\begin{equation} \label{R0}
R_0 := \frac{\beta p}{\mu_1} + \frac{\beta (1-p)}{\mu_2} < 1
\end{equation}
and so the left-hand side as a weighted average of reproduction numbers of both the symptomatic and asymptomatic cohorts can be defined as $R_0$. If $R_0 > 1$ the epidemic outbreaks. Since the infectious period $1/\mu_2$ of an Asymptomatic subject is much longer than the infectious period $1/\mu_1$ of a symptomatic Infectious subject, and $1-p$ is majority percentage for the COVID-19 disease, the derived $R_0$ is much higher than the one estimated using SEIR model (\cite{Kucharski}, \cite{Li}, \cite{Wu}). Violation of the condition \eqref{3} give birth to an unstable focus (two complex eigenvalues cross the imaginary axes). 

\subsection{Case Study: Numerical results for the COVID-19 Outbreak}

We do not aim here to fit and predict the curve of the infected. We are trying to explain that the principal difference between SEIR and SEIAR models leads to principally different outcomes in for example times and rates of outbreaks, a percentage of the population affected. It also has implications for state measures that are being introduced to keep the disease contained and controlled under health system capacity.

For commonly used parameters of COVID-19 $\beta=1$, $\gamma=1/4$, $\mu_1=1/3$, $\mu_2=1/10$ and $p=0.14$ (see \cite{Lauer}, \cite{Kucharski}, \cite{Li}, \cite{Wu}, \cite{NEJM}) we get $R_0 \doteq 9 > 1$ from \eqref{R0}. Condition \eqref{3} is satisfied. 
 
 \begin{figure}[h]
	\centering
	
	\includegraphics[width=0.53\textwidth]{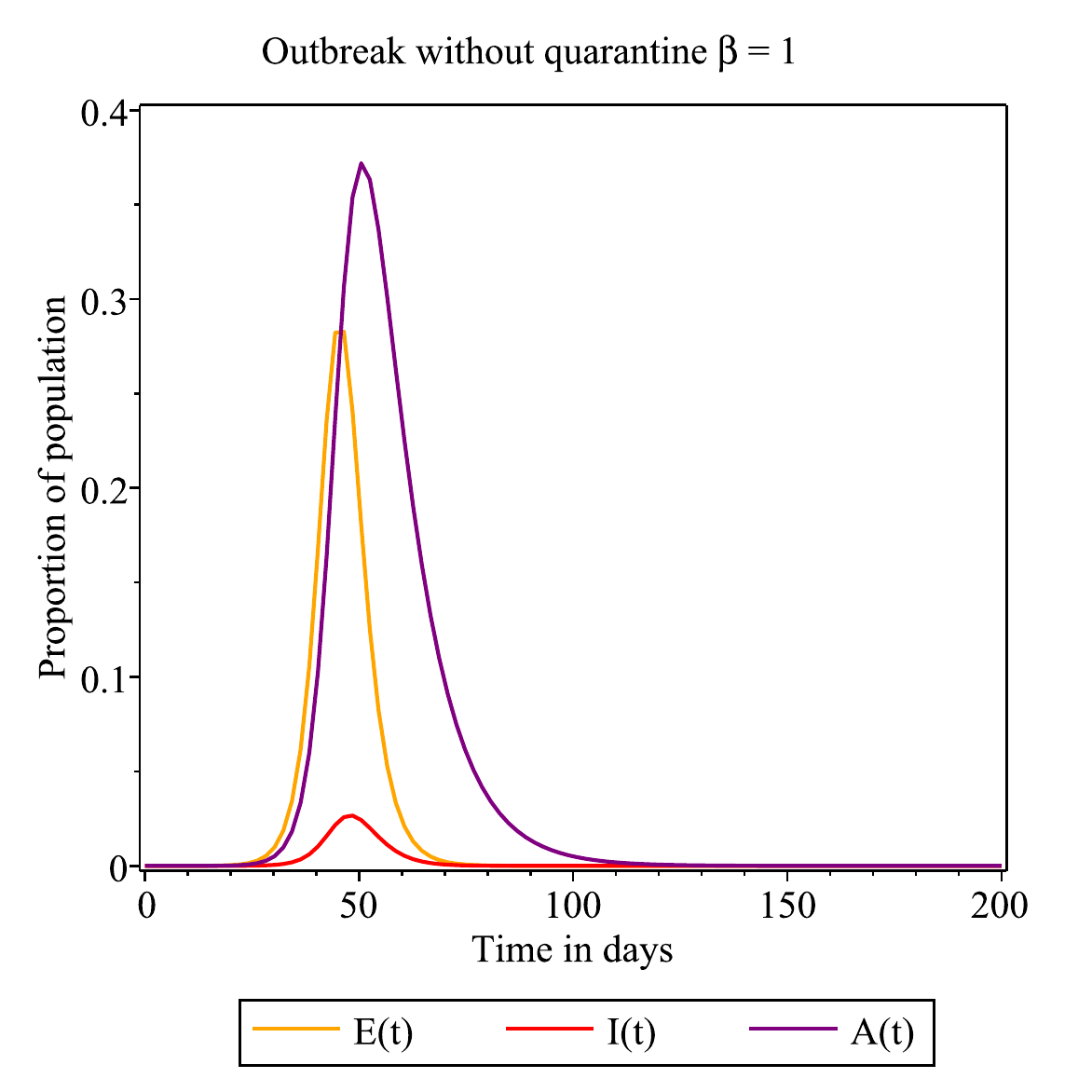}	
	
	\caption{Dynamics of the model  \eqref{S}, \eqref{E}, \eqref{I},\eqref{A}. The peak time $t_p=48$, a proportion of infectious in the population in the peak $I\left(48\right)=0.0268$.}
	\label{fig:SEIAdynamics}
\end{figure}

Figure \ref{fig:SEIAdynamics} shows the disease dynamics in case that no strong restrictions are held (without social distancing and other measures except hospitalization or isolated home treatment). The health system has to contain around $2.7 \%$ of Infected symptomatic population at once\footnote{This hypothetical case with no government interventions gives a peak around 280 000 people symptomatic altogether in Czech Republic.} at the moment of the peak (day 48 from the first Infected subject in a million population). Commonly, $20\%$ of symptomatic Infected subjects are hospitalized, and $10\%$ need intensive care (\cite{WHO}). That is $0.27 \%$ of the whole population. $60\%$ decrease of $\beta$ (e.g., mask protection in the whole population) causes the peak to be halved in approximately doubled time. SEIR model without the Asymptomatic cohort gives five times higher and about 3 weeks later peak values (for $\beta=1$), which seems to be very different from the actual dynamics of the disease COVID-19.

\subsection{GLEAMviz simulation of COVID-19 pandemic}

\begin{figure}[h]
	\centering
	\fbox{\includegraphics[width=\textwidth]{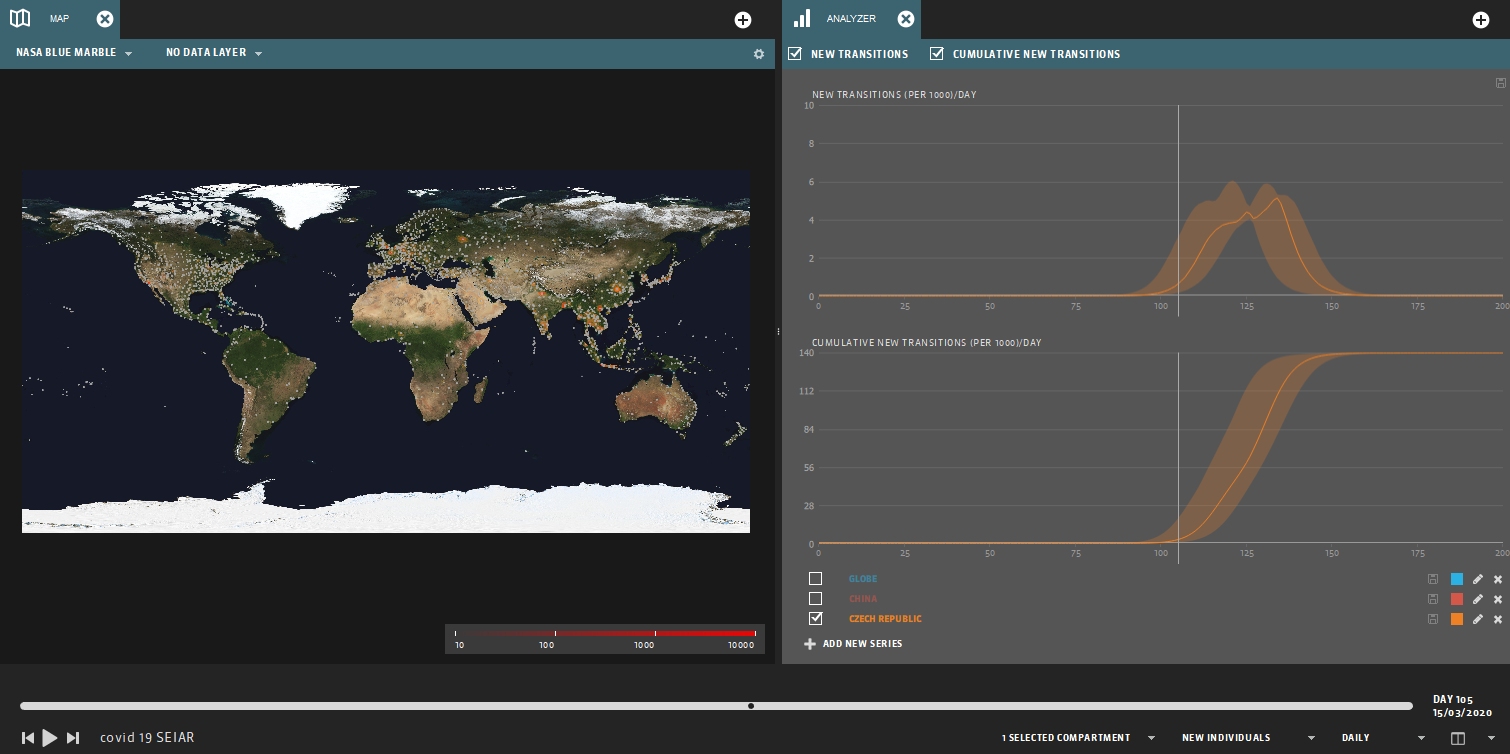}}
	\caption{Global simulation of COVID-19 outbreak without European protective measures in GLEAMviz based on the model  \eqref{S}, \eqref{E}, \eqref{I},\eqref{A} with parameters $\beta=1$, $\gamma=1/4$, $\mu_1=1/3$, $\mu_2=1/10$ and $p=0.14$. Czech Republic -- new transitions per 1000/day and cumulative transitions per 1000/day.}
	\label{fig:GLEAMviz}
\end{figure}

\begin{figure}[h]
	\centering
	\fbox{\includegraphics[width=\textwidth]{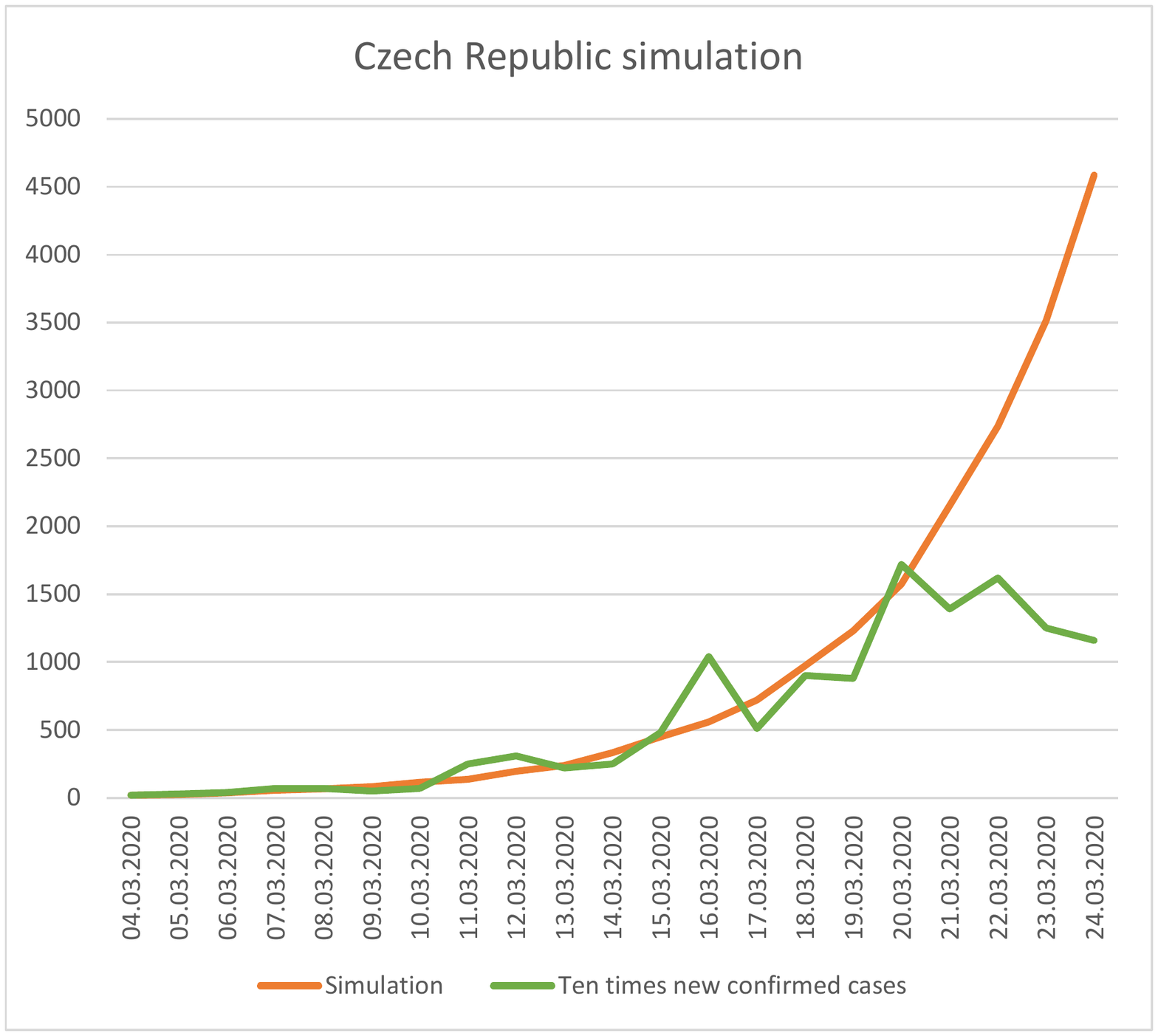}}
	\caption{COVID-19 outbreak without European protective measures in GLEAMviz based on the model  \eqref{S}, \eqref{E}, \eqref{I},\eqref{A} with parameters $\beta=1$, $\gamma=1/4$, $\mu_1=1/3$, $\mu_2=1/10$ and $p=0.14$ -- a comparison for Czech Republic new daily confirmed cases data set (\cite{GitHub}). Transmission rate $\beta$ is decreased to $0.3$ from 20/1/2020 for the whole China. Air traffic and local traffic is not decreased, there is no decrease of $R_0$ due to government measures in Europe, so the simulation fails after 20/03/2020 (around a week after first measures). Real data are 10 times lower (that could be due to the lack of testing at the beginning of the outbreak and undetected mild symptomatic cases) and delayed 16 days (due to the incubation period, developing symptoms period and testing period, plus possible time for undetected first cases). Simulation for SEIR model with the same parameters $(p=1, \, R_0=3)$ moves the outbreak more than a month later, and the infected cohort is more than five times bigger. This is evidence of the unsuitability of models without the asymptomatic infectious cohort since the parameters are based on clinical characteristics of the disease (commonly used, \cite{Lauer}, \cite{Kucharski}, \cite{Li}, \cite{Wu}, \cite{NEJM}).}
	\label{fig:CR}
\end{figure}

The \href{http://www.gleamviz.org/}{GLEAMviz simulation software} is a platform that gives a possibility to simulate the global pandemic on a network background. The Global Epidemic and Mobility Model (GLEAM) is a stochastic computational model that integrates high-resolution demographic and mobility data and uses a compartmental approach to define the epidemic characteristics of the infectious disease \cite{GLEAMviz}. We simulated the pandemic starting in Wu-Han on the 1st of December using SEIAR model with parameters $\beta=1$, $\gamma=1/4$, $\mu_1=1/3$, $\mu_2=1/10$ and $p=0.14$ and an exception that $\beta=0.3$ from 20/01/2020 in the whole China (Figure \ref{fig:GLEAMviz}). The outbreaks in Europe fit in time, unlike the much later onsets of the epidemics in case of using the SEIR model. Simulation-based on SEIAR showed good times of local epidemic outbreaks in Wu-Han, Italy, Germany, Czech Republic, and others (with differences in days,  Figure \ref{fig:CR}) and much closer proportion of infected in a population (difference in the order). Due to the gradual cancellation of flights, air traffic slowed down, so we cannot simulate the real situation. Possibility to decrease air traffic during simulations could be a good software improvement. 

\section{Efficiency of quarantine government measures with respect to the time of their introduction}

One important information that governments need is to estimate the efficiency of introduced measures. For the SEIAR model, we will show the dependence of the peak height on the time of the introduction of quarantine and social measures that affect transmission rate $\beta$. It turns out that there is a time threshold for the introduction of quarantine and social measures for both SEIR and SEIAR models\footnote{Model SEIR is a limit model SEIAR for $p=1$.}. Early measures lead to a significant reduction of the epidemic peak, whereas after a certain threshold, the epidemic peak cannot be significantly affected.

\begin{figure}[h]
	\centering
	\includegraphics[width=\textwidth]{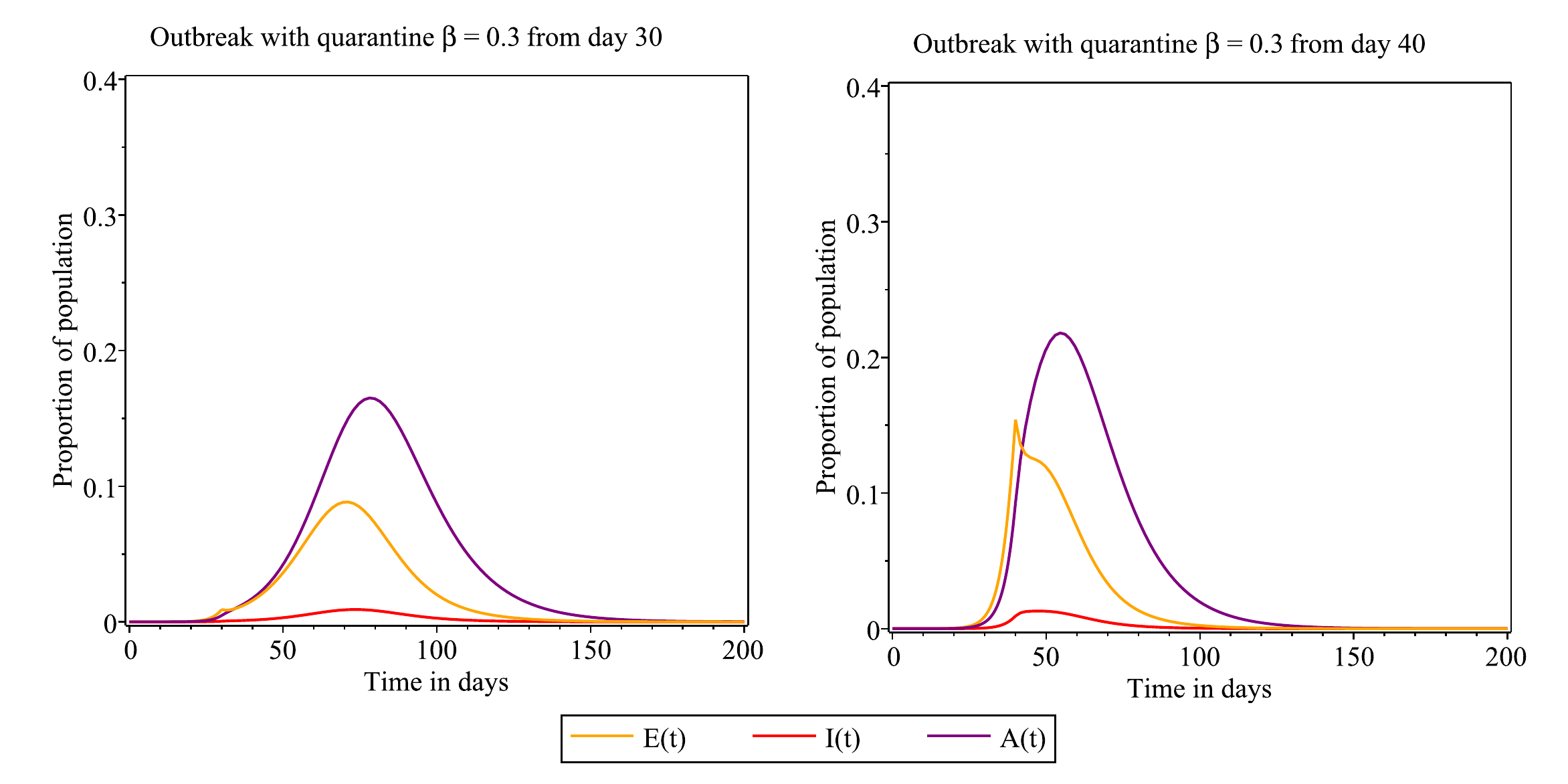}
	\caption{$\beta$ reduction from 1 to 0.3 is introduced at times $t=30$ and $t=40$. For $t=30$ the peak time $t_p=73$, a proportion of infectious in the population in the peak $I\left(48\right)=0.0091$. For $t=40$ the peak time $t_p=47$, a proportion of infectious in the population in the peak $I\left(48\right)=0.0130$.}
	\label{fig:jump}
\end{figure}

We will show this threshold for parameters $\beta=1$, $\gamma=1/4$, $\mu_1=1/3$, $\mu_2=1/10$ and $p=0.14$ used above . 
Figure \ref{fig:jump} shows how the dynamics of the system change abruptly as $\beta$ reduction from 1 to 0.3 is introduced at times $t=30$ and $t=40$. 
We have to highlight the fact that the time $t=30$ is counted from the first Infected subject in a million population, but due to delay (the incubation period, few more days of asymptomatic disease, test duration to confirm positivity) first COVID-19 patients in a new country were confirmed after around 10 days.

\subsection{Existence of a threshold}

\begin{figure}[h]
	\centering
	\includegraphics[width=0.56\textwidth]{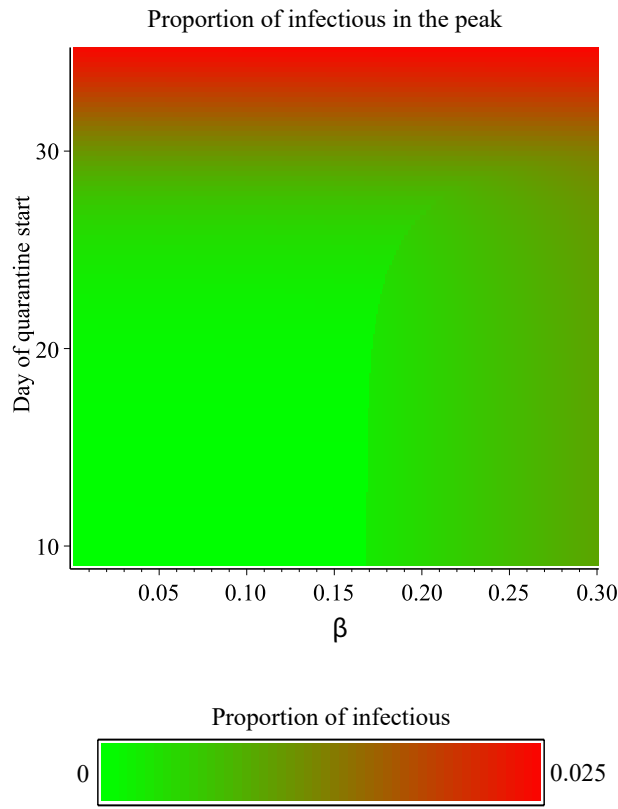}
	\caption{Proportion of infections in the peak depending on the strength of used measures (smaller $\beta$ means larger measures), and a number of days, when quarantine starts, from the first detected case (approximately 10 days after infection started). Days are shown from day 9, which is an estimation of a start of the phase 2 (outbreak). In case of two peaks of the epidemic, it displays the first peak.}
	\label{fig:heatmap}
\end{figure}

Depending on the time of the introduction of government measures that reduce $\beta$, we can draw the peak of the epidemic. The heatmap \ref{fig:heatmap} may serve to gain a good insight into the effectiveness of government measures in terms of time and strength. It displays a steep growth that is almost beta independent. This property is not exceptional for taken parameter values, but it is a generic behavior.

This could imply that countries that took action quickly (not later than 20-25 days from the first confirms) are probably on a good way to contain the outbreak. On the other hand, Italy, Spain, Switzerland, UK (with a lack of testing) or the US at highly populated areas face a sharp increase in the infected since the measures were introduced too late (this is of course not true for local areas with later outbreaks, where measures took place early enough).

\subsection{Efficiency threshold estimation}

It is challenging to estimate the efficiency of measures threshold during an ongoing outbreak. It is not possible to compare the time of the introduction of measures with the time of epidemic peak because we do not know it. A possible approach could be to compare the time of the fastest growth (denoted as Phase 2) with the time of introduction of measures.

\begin{figure}[h]
	\centering
	\fbox{\includegraphics[width=0.85\textwidth]{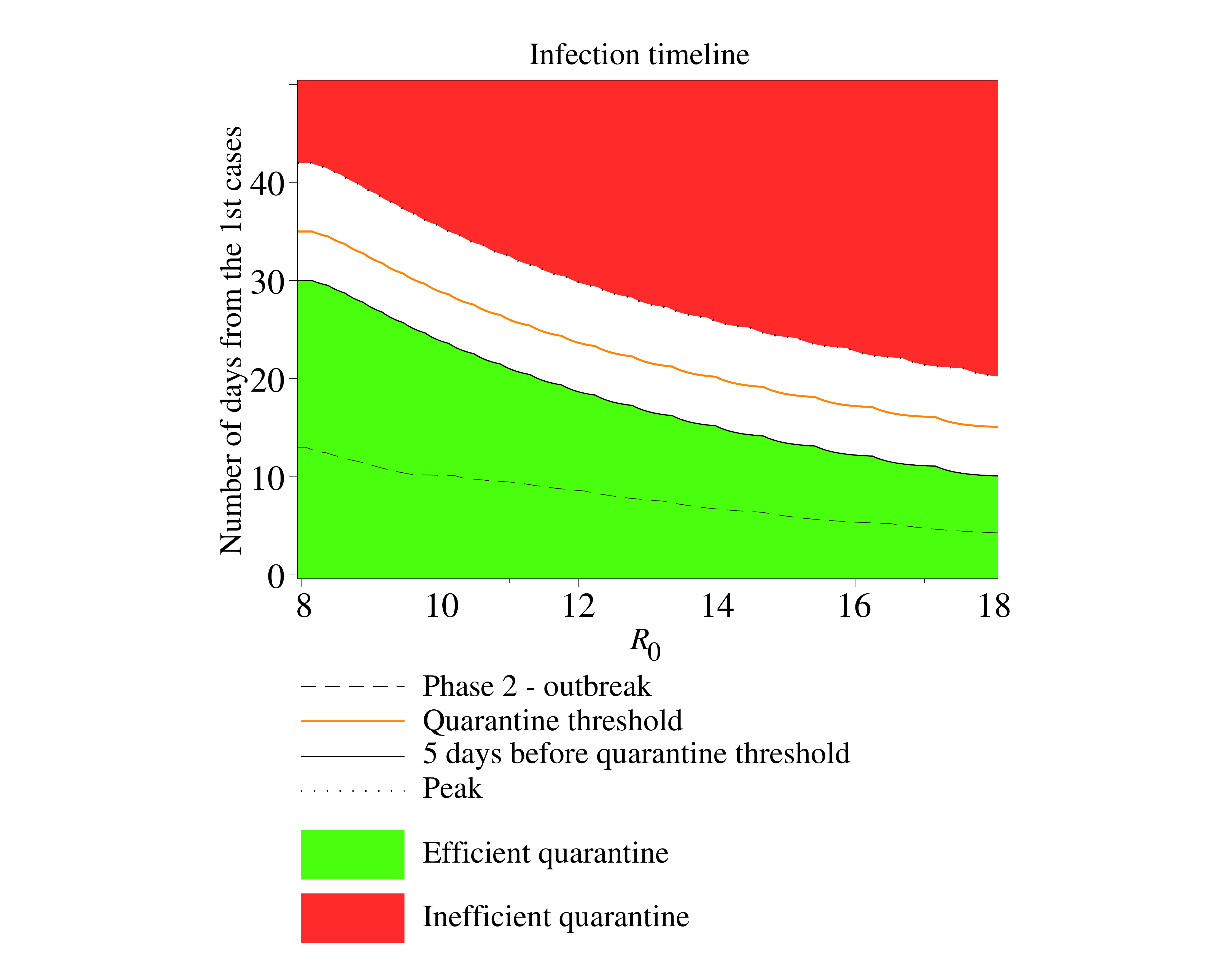}}
	\caption{Graph of quarantine efficiency for drop of $\beta$ to $0.3$ from a level given by $R_0$ ($\gamma=1/4$, $\mu_1=1/3$, $\mu_2=1/10$, $p=0.14$ $\beta$ given by \eqref{R0}). A period for efficient government measures is green -- from the 1st positive case to few days before the orange threshold. The dashed line indicates the outbreak of epidemic (maximal steep -- Phase 2). $R_0=9$ corresponds to clinical characteristics of COVID-19.}
	\label{fig:efekt}
\end{figure}

\begin{figure}[h]
	\centering
	\includegraphics[width=0.7\textwidth]{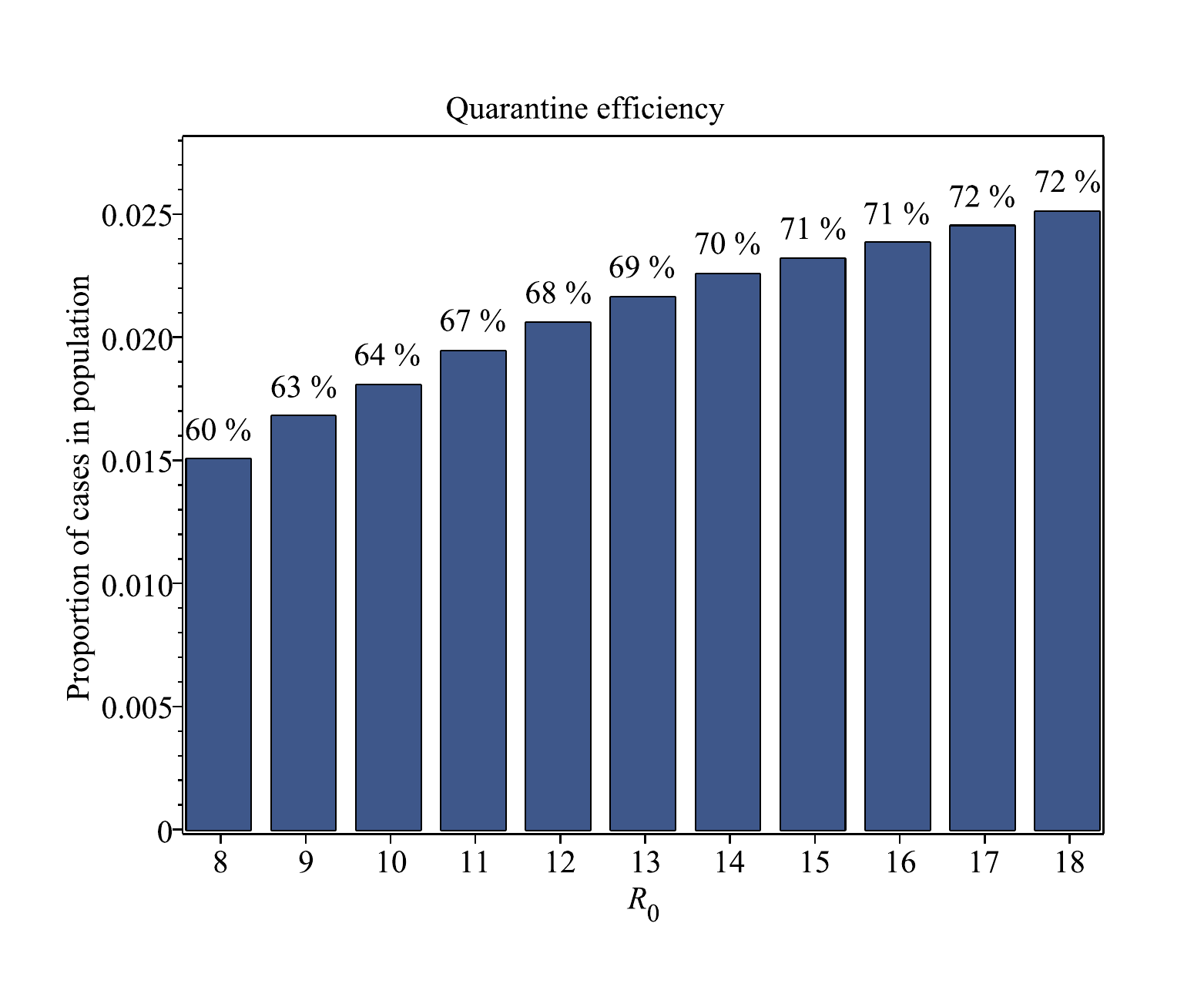}
	\caption{A difference between number of cases in the peak of the epidemic without a quarantine with $\gamma=1/4$, $\mu_1=1/3$, $\mu_2=1/10$, $p=0.14$ and $R_0$ given by \eqref{R0}, and the epidemic with a quarantine with $\gamma=1/4$, $\mu_1=1/3$, $\mu_2=1/10$, $p=0.14$, $\beta=0.3$. The percentages show a relative difference with respect to the peak of the epidemic without a quarantine.}
	\label{fig:compare}
\end{figure}

Figure \ref{fig:efekt} shows the possibility to efficiently contain the outbreak by quarantine, social distancing, mask-wearing, and other government measures that decrease transmission rate $\beta$ for various original $R_0$ ($\beta$ respectively). The dashed line indicates the outbreak of the epidemic (the maximal steep), the dotted line is the peak time. The threshold is orange, but the measures have to be introduced a few days before it. Again, parameters  $\gamma=1/4$, $\mu_1=1/3$, $\mu_2=1/10$ with $p=0.14$ are used. Figure \ref{fig:compare} shows the quarantine efficiency as decrease of infected in peak (proportion of population) for various original $R_0$. Each column is labeled with a percentage that shows a relative difference with respect to the peak of the epidemic without a quarantine.
 
\section{Conclusion}

We presented a compartmental model of epidemic spread, which is a generalization of the SEIR model by adding an asymptomatic infectious cohort. The derived basic reproduction number is the weighted arithmetic mean of the basic reproduction numbers of the symptomatic and asymptomatic cohorts. In a limit case (for an empty asymptomatic cohort), the SEIAR model is the SEIR model with standard $R_0$. This model was developed based on studies and clinical characteristics of COVID-19, but can be used, for example, to model the dynamics of the H1N1 influenza epidemic and other human-to-human transmission diseases. 

We are convinced that model SEIAR is the most simple model that can be used to simulate the epidemic dynamics of COVID-19. 
The standard SEIR model without an asymptomatic cohort does not match existing data, while the SEIAR model does quite well. Moreover, the onset of epidemics in different European countries corresponds to the simulations in GLEAMviz. The extremely rapid outbreaks in Italy and other countries, the functionality of only extreme measures, and the failure of measures that tried to protect and isolate only vulnerable groups are indirect supports for the suitability of the SEIAR model. Voices in the academic world are drawing attention to the asymptomatic group of infectious subjects. We are convinced that the asymptomatic cohort plays a crucial role in the spread of the COVID-19 disease.

The graph \ref{fig:heatmap} indicates that all measures need to be taken early and vigorously to be effective. Late or insufficient measures have little effect on the outbreak. All measures have economic, social, logistical, and psychological consequences. Their balance is up to the authorities.

\section*{Acknowledgements}

This work was supported by grant Mathematical and Statistical Modeling 4 number MUNI/A/1418/2019.

\bibliographystyle{unsrtnat}
\bibliography{references}  %%% Remove comment to use the external .bib file (using bibtex).

\end{document}